\documentstyle[preprint,aps]{revtex}
\begin{document}

\draft

\title{Heavy Meson Decays into Light Resonances}
\author{R. Delbourgo and Dongsheng Liu}
\address{Department of Physics, University of Tasmania\\
         Hobart, AUSTRALIA 7005}

\date{25 March, 1994}

\maketitle

\begin{abstract}
We analyse the Lorentz structures of weak decay matrix elements between 
meson states of arbitrary spin. Simplifications arise in the 
transition amplitudes for a heavy meson decaying into the light one
via a Bethe-Salpeter approach which incorporates heavy quark symmetry. 
Phenomenological consequences of our results on several semileptonic, 
nonleptonic and flavour changing neutral current-induced decays 
of heavy flavoured mesons are derived and discussed.  
\end{abstract}

\section{Introduction}
Decays of the $b$-quark into light $u,d,s$-quarks offer ways of testing 
the Standard Model and probing new physics. Thus the $b\rightarrow u$ 
decays give a direct determination of the CKM matrix element $|V_{ub}|$
while the flavour-changing neutral current (FCNC) induced transition
$b\rightarrow s$ allows one to extract knowledge of the 
yet to be discovered $t$-quark, and is sensitive to new physics 
beyond Standard Model. Although the charmed channels $b\rightarrow c$
dominate the total decay rate of $B$-meson, the much rarer decays of the 
$b$-quark into light flavoured quarks provide important information about 
the parameters of the CKM  matrix; in those rare decays there is so much
available phase space that it is possible to produce many light meson 
resonances (p, d, f...-wave) in the final state, not just the ground 
state mesons. Thus those rare processes tell us something about
the hadronic structure of light meson resonances, apart from giving us 
information about weak interaction elements;
hadronic matrix elements in weak decays have certainly attracted much
theoretical attention since it is hard to calculate them directly 
from the first QCD principles. The main purpose of this paper is to 
explore and understand the Lorentz structures of transition matrix 
elements of the weak current between meson states of arbitrary spin.
Most of the recent progress in the heavy quark effective theory has been
concentrated in the area of heavy hadron to heavy hadron transitions
\cite{hqet,BSmb,HtoH,FALK,Fahem,Salam,DL}; but a few results have appeared
in the literature about heavy to light hadron transitions in
their ground states\cite{IW42,BSmb}. It is our aim here to extend the 
arguments to the heavy to light meson resonances of arbitrary spin.

We will begin by outlining a general rule for counting the number of 
form factors representing independent Lorentz structures in Section II
and we present explicit forms of those elements for processes, 
$0\rightarrow J^{\prime}$ and $1\rightarrow J^{\prime}$ before detailing 
the general case $J \rightarrow J'$. It is widely accepted that the 
heavy quark limit is a reliable approximation for treating mesons and 
baryons containing a $b$-quark, as long as the recoil between the $b$ and 
the light degrees of freedom is not particularly large. Therefore in 
Section III we take that limit for the initial $B$-mesons and show
how the number of form factors is reduced. The expressions for decay 
rates (in terms of those form factors) are worked out in Section IV
and our conclusions are stated in Section V. An appendix contains
some useful technicalities about sums over polarization tensors.
     
\section{ Hadronic Matrix Elements }

We begin with considering the  matrix elements of a vector (or axial) 
current between a 
spin $J=0$ meson state of momentum $p$ and a meson resonance of momentum
$p'$ with spin $J'$. For the simple case when $J'=0$, it is well-known
that there are just two Lorentz invariant form factors parametrizing the 
matrix element:
\begin{equation}
\langle p',0|{\cal J}_\mu|p,0 \rangle 
               \equiv f_1(p.p')p_\mu+f_2(p.p')p'_\mu
               \equiv a_+(q^2)(p+p')_\mu + a_-(q^2)(p-p')_\mu,
\end{equation}
where $q=p-p'$ and we have made no assumptions about current conservation
at this stage. Before we present the results for any $J'$, let us
take the $J'=1$ case as an example of the proliferation of form factors
when $J'$ and $J$ grow. Here, in addition to the vectors $p_\mu$ and 
$p'_\mu$, we have available the final vector meson polarization vector
 $\phi'^*_{\mu}(p')$, satisfying $\phi'_\mu p'^\mu=0$. Allowing also for
the Levi-Civita tensor (since we have made no assumptions about parity as 
yet), we arrive at four form factors:
$$\langle p',1|{\cal J}_\mu|p,0\rangle \equiv
  [a_+(q^2)(p+p')_\mu + a_-(q^2)(p-p')_\mu]p^\nu\phi'^*_\nu + 
  f(q^2)\phi'^*_\mu+ig(q^2)\epsilon_{\mu\alpha\beta\gamma}\phi'^{*\alpha}
                                  p^\beta p'^\gamma. $$
Naturally, if we impose parity conservation, then either one or
three of the above structures disappear.

For higher spin $J'$, we may represent the final meson by a Lorentz tensor 
of rank $J'$, namely $\phi'^*_{\{\mu_1\cdots\mu_{J'}\}}(p')$; it is 
of course transverse to $p'$, symmetric and traceless. Because the final
result must be a Lorentz vector, the indices of the polarization tensor
may either be completely saturated with $p$ to form the scalar 
$\phi'^*_{\{\mu_1\cdots\mu_{J'}\}} p^{\mu_1}\cdots p^{\mu_{J'}},$
or we may leave one index free, $\phi'^*_{\{\mu\mu_2\cdots\mu_{J'}\}}
 p^{\mu_2}\cdots p^{\mu_{J^{\prime}}}$. As well we should allow for a 
Levi-Civita tensor coupling to the polarization tensor in the form
$$ \epsilon_{\mu\mu_1\alpha\beta} 
   (\phi'^{*\{\mu_1\mu_2\cdots\mu_{J'} \}}
     p_{\mu_2}\cdots p_{\mu_{J'}}) p^\alpha p'^\beta. $$
Altogether then we can construct the vectorial matrix element
\begin{equation}\begin{array}{rl}
 \langle p',J' \geq 1|{\cal J}_\mu|p,0\rangle
 & = a_+^{(J')}(q^2) (\phi'^*_{\{\mu_1\cdots\mu_{J'}\}} 
                   p^{\mu_1}\cdots p^{\mu_{J'}}) (p+p')_\mu \\
 \displaystyle
 & + a_-^{(J')}(q^2) (\phi'^*_{\{\mu_1\cdots\mu_{J'}\}}
                   p^{\mu_1}\cdots p^{\mu_{J'}}) (p-p')_\mu \\
 \displaystyle
 & + f^{(J')}(q^2) (\phi'^*_{\{\mu\mu_2\cdots\mu_{J'}\}}
                  p^{\mu_2}\cdots p^{\mu_{J'}}) \\
 \displaystyle
 & + ig^{(J')}(q^2) \epsilon_{\mu\mu_1\alpha\beta} 
     (\phi'^{*\{ \mu_1\mu_2\cdots\mu_{J'}\}}
     p_{\mu_2}\cdots p_{\mu_{J^{\prime}}}) p^\alpha p'^\beta.
\end{array}\end{equation}

Some remarks are in order at this stage:

\noindent
(1) Except for the case $J'=0$ case where the number of independent 
    form factors is two, there are at most {\em four} form factors for
    arbitrary higher spin $J'$ when $J=0$.

\noindent
(2) The number of form factors depends of course on the angular 
    momentum which the current carries. The transverse part of a vector
    (or axial) current will carry spin-$1$, while the longitudinal 
    part corresponds to spin-$0$ and is not relevant for conserved
    currents. Thus we may associate three pieces of (2) with orbital
    angular momentum $J'+1, \; J'$ and $J'-1$ for the transverse current, 
    and one orbital piece $J'$ for the longitudinal current. 
    In the special case $J'=0$, one transverse form factor and one 
    longitudinal form factor survive.   

\noindent
(3) Keeping these points in mind and following the authors of ref. 
   \cite{WSB}, it is appropriate to express the vectorial matrix element 
   for general $J'$ in the form,
   \begin{equation}\begin{array}{rl}
    \langle p',J'|{\cal J}_\mu|p,0 \rangle
    \displaystyle
    & = F^{(J')}_0(q^2)(\phi'^*_{\{\mu_1\cdots\mu_{J'}\}}
                  p^{\mu_1}\cdots p^{\mu_{J'}})
    \frac{M^2-M'^2}{q^2}q_\mu \\
    \displaystyle
    & + F_1^{(J')}(q^2)(\phi'^*_{\{\mu_1\cdots\mu_{J'}\}}
                  p^{\mu_1}\cdots p^{\mu_{J'}})
      [(p+p')_\mu - \frac{M^2-M'^2}{q^2}q_\mu ], \\
   \displaystyle
   & +F_2^{(J')}(q^2)
   \left [\phi'^*_{\{\mu\mu_2\cdots\mu_{J'}\}}
      p^{\mu_2}\cdots p^{\mu_{J'}} -\frac{q_\mu}{q^2}
      \phi'^*_{\{\mu_1\cdots\mu_{J'}\}} p^{\mu_1}\cdots p^{\mu_{J'}}
           \right ] \\
   \displaystyle
   & + F_3^{(J')}(q^2) \epsilon_{\mu\alpha\beta\gamma} 
    \phi'^{*\{\alpha\mu_2\cdots\mu_{J'}\}}
     p_{\mu_2}\cdots p_{\mu_{J'}} p^\beta p'^\gamma,
  \end{array}\end{equation}
  where the last two terms are not present when $J'=0$.

With the experience gained from the work presented in previous paragraph, 
we may now analyse matrix elements between initial spin-$1$ state
and a final state of arbitary spin. For the $1\rightarrow 0$ case,
all results for $0\rightarrow 1$ discussed previously are retained except
that one should replace the final polarization vector by the initial 
one and exchange momenta (crossing). 

When both final and initial spins are 1, we have two polarization vectors, 
the initial $\phi_\mu$ and the final $\phi'^*_\nu$; the matrix element 
is a bilinear of them. From them we can firstly construct scalar 
invariants like
$$(\phi_\mu p'^\mu)(\phi'^*_\nu p^\nu), \quad
  (\phi_\mu\phi'^{*\mu}), \quad
  \epsilon^{\alpha\beta\gamma\delta}
\phi_\alpha \phi'^*_\beta p_\gamma p'_\delta. $$
In combination with either $p_\mu$ or $p'_\mu$, each of the above three
scalars generates two form factors; so at the moment we have six in all.
Next, allowing the polarization vectors to carry the Lorentz index 
of the current, we gain two more form factors:
$$\phi_\mu(\phi'^*_\nu p^\nu), \quad (\phi_\nu p'^\nu)\phi'^*_\mu. $$
Finally, using the Levi-Civita tensor we get two more structures,
$$\epsilon_{\mu\alpha\beta\gamma}\phi^\alpha\phi'^{*\beta}p^\gamma \quad
  {\rm and}\quad \epsilon_{\mu\alpha\beta\gamma}
   \phi^\alpha\phi'^{*\beta}p'^\gamma. $$
In all there are therefore {\em ten} form factors in the case $J=J'=1$. 
Although one can contemplate structures like
$$(\phi^{\prime}_{\nu}p^{\nu})\epsilon^{\mu\alpha\beta\gamma}
  \phi_{\alpha}p_{\beta}p'_{\gamma}, \quad
  (\phi_\nu p'^\nu)\epsilon^{\mu\alpha\beta\gamma}
  \phi'_\alpha p_\beta p'_\gamma,$$
the identity
$$P_{\mu}\epsilon_{\alpha\beta\gamma\lambda}
 =P_{\alpha}\epsilon_{\mu\beta\gamma\lambda}
 +P_{\beta}\epsilon_{\alpha\mu\gamma\lambda}
+P_{\gamma}\epsilon_{\alpha\beta\mu\lambda}
+P_{\lambda}\epsilon_{\alpha\beta\gamma\mu},$$
can be massaged to show that these new terms are not independent of the 
previous ones.

Now we proceed to matrix elements for $1\rightarrow J^{\prime}\geq 2$.
Besides the polarization vector $\phi_\mu$ of the initial meson, we have 
the Lorenz tensor of rank $J'$, $\phi'_{\{\mu_1\cdots\mu_{J'}\}}$
for the final meson. When this polarization tensor occurs in the 
contracted vector form
$$ \varphi'_{\mu}
  =\phi'_{ \{ \mu\mu_2\cdots\mu_{J'}\}}p^{\mu_2}\cdots p^{\mu_{J'}}, $$
we can repeat the earlier analysis ($1\rightarrow 1$) and obtain
ten form factors. In addition we should consider the possibility that
two indices remain uncontracted:
$$ \varphi'_{\mu\nu}=\phi'_{\{\mu\nu\mu_3\cdots\mu_{J'}\}}
   p^{\mu_3}\cdots p^{\mu_{J'}}, $$
from which we can build two more vectorial covariants,
$\phi^\nu\varphi'_{\mu\nu}$ and $\epsilon^{\mu\alpha\beta\gamma}
\phi^\nu\varphi'_{\alpha\nu}p_\beta p'_\gamma.$ 
Hence the number of form factors rises to {\em twelve}.
As far as the counting is concerned, this ties in very nicely with the 
classical analysis based on angular momentum addition:

\noindent
(1) Letting $S=1$ correspond to the transverse current and coupling it
    to $J=1$, we obtain total spin $2,1,0$. To these we may add orbital
    angular momenta $L=J'+2,J'+1,J',J'-1,J'-2$, $L=J'+1,J',J'-1$ and
    $L=J'$, respectively. Consequently, there are $5+3+1 = 9$ form factors 
    when $J'\geq 2$. 

\noindent
(2) Setting $S=0$ for the longitudinal part, there is only total spin 1
    (of the initial meson). So here we get 3 form factors, associated 
     with $L = J'+1 , J', J'-1$, provided that $J'\geq 1$. 

\noindent
Adding (1) and (2), the total (maximum) number of form factors is twelve 
--- but is of course reduced to a fewer number when $J'\leq 1$.

We are now in a position to outline the general rule for counting how 
many independent form factors are needed to describe a vector (or axial) 
current between spin-$J$ and spin-$J'$ meson states. The analysis
is best carried out in the channel of the current. First we decompose the 
current itself into a transverse part ($S=1$) and a longitudinal part 
($S=0$). Secondly we compose the spins of the two mesons into the set
$$J+J',\;J+J'-1,\ldots,|J-J'|$$
and ask what angular momentum values $L$ are needed to give total spin $S$.
For $S=0$, $L$ necessarily equals the total mesons' spin, while 
for $S=1$ there is a three-fold possibility for $L$ (assuming the total 
mesons' spin exceeds 0). Hence the total number of $L$-values, and thus 
form factors, equals the sum $N$ of $N_0$ and $N_1$ where
$$ N_0 = 1 + 2 {\rm Min}(J,J');\qquad S=0$$
$$ N_1 = 1 + \sum_{k=0}^{2{\rm Min}(J,J')}[1+2{\rm Min}(1,|J-J'|+k) + 1];
   \qquad S=1,$$
leading to
$$ N = 4(2J'+1); \qquad {\rm for~}J'<J,$$
$$ N = 4(2J+1) - 2; \qquad {\rm for~} J'=J,$$
$$ N = 4(2J+1);\qquad {\rm for~} J'>J.$$ 
These structures may be given a Lorentz covariant form. We shall not
write them all out as they are not needed in the present investigation. 
We will just content ourselves by stating what they reduce to 
at a special kinematical point, zero recoil, where $Mp'=M'p$ and $M,M'$ 
stand for the masses of the mesons in the initial and final states. 
In this limit only 3 structures survive for the element 
$\langle J'|{\cal J}_\lambda|J\rangle$, namely, 
$$\phi'^{*\{\mu_1\cdots\mu_{J'}\}}\phi_{\{\lambda\mu_1\cdots\mu_{J'}\}};
  \qquad J'=J-1, $$
$$\phi'^{*\{\mu_1\cdots\mu_{J}\}}\phi_{\{\mu_1\cdots\mu_{J}\}}
    (p+ p')_\lambda;  \qquad J'=J, $$
$$\phi'^*_{\{\lambda\mu_1\cdots\mu_{J}\}}\phi^{\{\mu_1\cdots\mu_{J}\}};
 \qquad J'=J+1. $$

\section{ Heavy to Light Transitions }

It has been demonstrated that Bethe-Salpeter approach is as useful as 
the so-called tensor method of the heavy quark effective theory for 
treating hadronic matrix elements \cite{BSmb,Fahem,DL}. We shall use 
this approach, incorporating heavy quark symmetry, to investigate weak 
decays of a heavy meson into light resonances in this section.
In analogy to the interpolating field method, we shall consider
the Bethe-Salpeter amplitude $\phi_{\alpha}^{\beta}
=<0|T Q_{\alpha}\bar q^{\beta} |p>$ for the meson state in 
momentum-space, 
\begin{equation}
 \phi_{\alpha}^{\beta}
=[\chi_{ ( \mu_1\cdots\mu_L ) }(p)]^{\sigma}_{\rho}
\;( A^{\mu_1\cdots\mu_L } )_{\sigma\alpha}^{\rho\beta},
\end{equation}
where the spin-parity projector $\chi$ represents the Lorentz-covariant
wavefunction of the external meson. The structure of spin-parity projectors
for resonances of higher spin have been worked out by us \cite{Salam,DL} 
and here we just list the results: 
\begin{equation}
\chi_{ ( \mu_1\cdots\mu_L ) }^{(^1 L_L)}(p)
=\gamma_5 P_{ \{ \mu_1\cdots\mu_L \} }(p),
\end{equation}
\begin{equation}
\chi_{ ( \mu_1\cdots\mu_L ) }^{(^3L_{L+1})}(p)=\gamma^{\mu} 
V^{L+1}_{ \{ \mu\mu_1\cdots\mu_L \} }(p),
\end{equation}
\begin{equation}\displaystyle
\chi_{ ( \mu_1\cdots\mu_L ) }^{(^3L_L)}(p)=-i\gamma^{\mu} 
\sum_k\frac{p^{\lambda}}{m}\epsilon_{\lambda\mu\mu_k\nu}d^{\nu\nu'}(p)
V^{L}_{ \{ \mu_1\cdots\bar k\cdots\mu_L\nu' \} }(p),
\end{equation}
and
\begin{equation}\displaystyle
\chi_{ ( \mu_1\cdots\mu_L ) }^{(^3L_{L-1})}(p)=\gamma^{\mu} 
\left [ 
\sum_k d_{\mu\mu_k}(p)
V^{L-1}_{ \{ \mu_1\cdots\bar k\cdots\mu_L \} }(p)
\displaystyle
-\frac{2}{2L-1} \sum_{kl} d_{\mu_k\mu_l}(p)
  V^{L-1}_{ \{ \mu\mu_1\cdots\bar k\bar l\cdots\mu_L \} }(p)
\right ],
\end{equation}
where we have adopted the standard notation $^{2S+1}L_J$ and 
$d_{\mu\nu}(p)$ is given in the Appendix. 
The parity of the meson resonances is given by $(-1)^{L+1}$ and $CP=-1$
for the singlet and $+1$ for triplet. 
When a heavy meson contains an on-shell heavy quark, one has further
\begin{equation}\displaystyle
 \phi_{\alpha}^{\beta}
=[\frac{1+\not v}{2}\chi_{ ( \mu_1\cdots\mu_L ) }(v)]^{\sigma}_{\alpha}
\;( A^{\mu_1\cdots\mu_L} )_{\sigma}^{\beta}.
\end{equation}

For heavy mesons, it is conventional to organize the terms as 
eigenfunctions of projectors corresponding to the total angular momentum 
of the light degrees of freedom.  Even though we really have no detailed 
knowledge of the configuration of the light degrees of freedom, the 
decoupling of the heavy quark spin tells us the two components in a 
doublet generated by the heavy quark spin operator tie in with those of 
the light degrees of freedom. Using this line of argument the spin-parity 
operators have been presented by Falk \cite{FALK} and are related to ours 
through Clebsch-Gordan coefficients. We will return to this issue soon. 
The matrix element for the heavy to light transition takes the 
form \cite{fn1}

\begin{equation}\displaystyle
<X_L(p)|\bar q \Gamma h_v |X(v)>=
{\rm Tr}\left [ M^{\nu_1\cdots\nu_L}(v,p) \Gamma 
          \frac{1+\not v}{2}\chi_{ ( \nu_1\cdots\nu_L ) }(v) \right ]. 
\end{equation}
The overlap integral $M$ involves the light degrees of freedom 
in both heavy and light hadrons and embodies the spin and parity of 
the light meson, 
\begin{equation}
 (M^{\nu_1\cdots\nu_L})_{\alpha}^{\beta}
 =[\bar \chi_{ ( \mu_1\cdots\mu_{L'} ) }(p)]^{\sigma}_{\rho}\;
  [{\cal M}^{\mu_1\cdots\mu_{L'};
           \nu_1\cdots\nu_L }(p, v)]_{\sigma\alpha}^{\rho\beta}.
\end{equation}
Here ${\cal M}^{\mu_1\cdots\mu_L;\nu_1\cdots\nu_L }$ is a 
Lorentz tensor with parity $(-1)^{L+L'}$, which contains all the
non-perturbative physics of the matrix element,  carrying also the 
symmetry properties of the external current $\Gamma$.
However, as far as the multispinor is concerned, we can always decompose 
${\cal M}$ into ${\cal M}={\cal D}\otimes D$, with $D$ being
one of 
$\Gamma=I,\gamma_5,\gamma_{\lambda},\gamma_{\lambda}\gamma_5, 
\sigma_{\lambda\gamma}$, and attribute all momentum dependence to 
${\cal D}$. Thus we can always rewrite the overlap integral in the form
\begin{equation}
M_{\alpha}^{\beta}={\cal D}_{\alpha}^{\beta}{\rm Tr}[D\bar \chi(p)].
\end{equation}
Evidently for $S=0$ resonances, only the contribution 
corresponding to $D=\gamma_5$ to the overlap integral $M$ survives, whilst
the part $D=\gamma_{\lambda}$ is associated with $S=1$.
Based on this, we shall construct the most general form of $M$ in terms of 
the tensors $P$ and $V$, which are symmetric, transverse and traceless.

(1) $s$-wave $\rightarrow J'$ 

Firstly let us examine decays of a heavy meson of the $(0^-,\; 1^-)$ 
doublet into a light resonance of higher spin. The spin-parity projector 
for the heavy meson is spin-$0$, {\sl i.e.} $\chi = \gamma_5$ and
$\gamma\cdot V$, so we need only find the general form for the overlap 
integral with no Lorentz index. For a final resonance which is a spin 
singlet, we remain with 
\begin{equation}\displaystyle
\begin{array}{rl}
 M^{(^1L'_{L'})}(v,p)=&\gamma_5\left[ G_1^{(^1L'_{L'})}(v\cdot p)
       +G_2^{(^1L'_{L'})}( v\cdot p)\not p\right ]
      v^{\mu_1}\cdots v^{\mu_{L'}} P_{ \{ \mu_1\cdots\mu_{L'}\} }(p) \\
       + & \gamma_5 
\left [ G_3^{(^1L^{\prime}_{L'})}( v\cdot p)
      + G_4^{(^1L^{\prime}_{L'})}( v\cdot p)\not p \right ]
 \gamma^{\mu_1}v^{\mu_2}\cdots v^{\mu_{L'}}P_{\{\mu_1\cdots\mu_{L'}\}}(p).
\end{array}
\end{equation}
To explain why that is all, we note that 
$P_{\{\mu_1\cdots\mu_{L'}\}}$ is transverse to $p$ and traceless so that 
only products of $v^{\mu}$ and $\gamma^{\mu}$ may be contracted with it. 
Hence a product involving $v^{\mu}$ purely gives the first two form 
factors in Eq.(13). Furthermore, since 
$\gamma^\mu\gamma^\nu=g^{\mu\nu}-i\sigma^{\mu\nu},$ and bearing in mind
the symmetry property, two gamma matrices are not permitted; only one 
$\gamma^{\mu}$ is allowed, its position being irrelevant.
Hence we have the second two terms (which are absent when $L'=0$).

We turn now to the spin triplet. For $J'=L'+1$ resonances, we shall build 
up the Dirac bispinor $M$ using $V^{L'+1}_{\{\mu\mu_1\cdots\mu_{L'}\}}(p)$ 
and $v$ and Dirac matrices. Notice that $V^{L'+1}$ has the same properties
as $P$ in Eq.(13), except for the rank which does not matter. 
This leads us to
\begin{equation}\displaystyle
\begin{array}{rl}
 M^{(^3L^{\prime}_{L'+1})}(v,p)=&   
\left [ G_1^{(^3L'_{L'+1})}( v\cdot p) 
      + G_2^{(^3L'_{L'+1})}( v\cdot p)\not p \right]v^\mu 
    v^{\mu_1}\cdots v^{\mu_{L'}}V^{L'+1}_{\{\mu\mu_1\cdots\mu_{L'}\}}(p)
 \\
       + & 
\left [ G_3^{(^3L'_{L'+1})}( v\cdot p)
      + G_4^{(^3L'_{L'+1})}( v\cdot p)\not p \right]
      \gamma^{\mu}v^{\mu_1}\cdots v^{\mu_{L'}} 
      V^{L'+1}_{ \{ \mu \mu_1\cdots\mu_{L'} \} }(p).
\end{array}
\end{equation}
The analysis for $J'=L'$ resonances of the spin triplet instead goes as
follows. On the face of it we can construct the following Lorentz scalars
$$\displaystyle
\gamma^{\mu} \sum_k
\frac{p^\lambda}{m} v^{\mu_k}\epsilon_{\lambda\mu\mu_k\nu}d^{\nu\nu'}
V^{L'}_{\{\mu_1\cdots\bar k\cdots\mu_{L'}\nu' \} }(p)
\underbrace{ v^{\mu_1}\cdots v^{\mu_{L'}} }_{ without\; v^{\mu_k} },$$
$$\displaystyle
\gamma^{\mu} \sum_k
\frac{p^\lambda}{m}\gamma^{\mu_k}\epsilon_{\lambda\mu\mu_k\nu}d^{\nu\nu'}
 V^{L'}_{ \{ \mu_1\cdots\bar k\cdots\mu_{L'}\nu' \} }(p)
 \underbrace{ v^{\mu_1}\cdots v^{\mu_{L'}} }_{ without\; v^{\mu_k} },$$
$$\displaystyle
\gamma^{\mu} \sum_k
\frac{p^\lambda}{m} v^{\mu_k}\epsilon_{\lambda\mu\mu_k\nu}d^{\nu\nu'}
 V^{L'}_{ \{ \mu_1\cdots\bar k\cdots\mu_{L'}\nu' \} }(p) \gamma^{\mu_1}
\underbrace{ v^{\mu_2}\cdots v^{\mu_{L'}} }_{ without\; v^{\mu_k} },$$
$$\displaystyle
\gamma^{\mu} \sum_k
\frac{p^\lambda}{m}\gamma^{\mu_k}\epsilon_{\lambda\mu\mu_k\nu}d^{\nu\nu'}
 V^{L'}_{\{ \mu_1\cdots\bar k\cdots\mu_{L'}\nu' \} }(p) \gamma^{\mu_1}
 \underbrace{ v^{\mu_2}\cdots v^{\mu_{L'}} }_{ without\; v^{\mu_k}},$$
and terms with an additional $\not p$. However using identities
$$ i \gamma^{\mu} \epsilon_{\mu\nu\lambda\sigma}
  =\gamma_5( g_{\nu\lambda}\gamma_{\sigma}
           +g_{\lambda\sigma}\gamma_{\nu}
           -g_{\sigma\nu}\gamma_{\lambda}
           -\gamma_{\nu}\gamma_{\lambda}\gamma_{\sigma} ),$$
and
$$ i \sigma^{\mu\nu} \epsilon_{\mu\nu\lambda\sigma}
=2\gamma_5\sigma_{\lambda\sigma},$$ 
all of these structures can be reduced into four independent forms, namely
\begin{equation}\displaystyle
\begin{array}{rl}
 M^{(^3L'_{L'})}(v,p)=&\gamma_5   
 \left[ G_1^{(^3L'_{L'})}( v\cdot p) 
      + G_2^{(^3L'_{L'})}( v\cdot p)\not p \right]
  V^{L'}_{ \{ \mu_2\cdots\mu_{L'}\nu' \} }(p)
           v^{\mu_2}\cdots v^{\mu_{L'}} v^{\nu'} \\
       +&\gamma_5   
\left [ G_3^{(^3L'_{L'})}( v\cdot p) 
      + G_4^{(^3L'_{L'})}( v\cdot p)\not p \right]
  V^{L'}_{ \{ \mu_2\cdots\mu_{L'}\nu' \} }(p)
           v^{\mu_2}\cdots v^{\mu_{L'}} \gamma^{\nu'}
\end{array}
\end{equation}
Apart from the change in the value of $S$ the structure of $M(v,p)$
is exactly the same as that for resonances of spin singlet in Eq.(13).
Following a similar procedure it is straightforward to work
out the results for spin triplet of $J'=L'-1$; these read
\begin{equation}\displaystyle
\begin{array}{rl}
 M^{(^3L'_{L'-1})}(v,p)=&   
\left [  G_1^{(^3L'_{L'-1})}( v\cdot p) 
       + G_2^{(^3L'_{L'-1})}( v\cdot p)\not p \right]
 v^{\mu_2}\cdots v^{\mu_{L'}} V^{L'-1}_{ \{ \mu_2\cdots\mu_{L'}\}}(p)\\
       + & 
\left [ G_3^{(^3L'_{L'-1})}( v\cdot p)
      + G_4^{(^3L'_{L'-1})}( v\cdot p)\not p \right]
      \gamma^{\mu_2}v^{\mu_3}\cdots v^{\mu_{L'}}
       V^{L'-1}_{ \{ \mu_2\mu_3\cdots\mu_{L'} \} }(p) ,
\end{array}
\end{equation}
in which only the first two form factors contribute when $L'=1$. 

Compared with the general problem, discussed in Section II, 
the simplification resulting from the heavy quark approximation 
is two-fold: in the first place the decaying $1^-$ meson shares the same 
complexity as a $0^-$ meson; given the state of the light degrees of 
freedom of the $(0^-, 1^-)$ doublet, the overlap integral is actually 
determined by the state of the light resonance; in the second place
a set of four `universal' form factors are sufficient to parametrize all 
matrix elements of bilinear operators $ \bar q \Gamma h_v$ 
for each $^{2S+1}L_J$-configuration. However given a particular current 
it is possible for some of them to be absent.

(2) $p$-wave $\rightarrow J'$

Here it will prove convenient to mix the $^1P_1$ and $^3P_1$ states of 
the heavy decaying meson to track the spin of the constituent light 
degrees of freedom. The way to do this has been delineated in
ref.\cite{FALK}, and in our case we state the decomposition much more 
explicitly. Given our spin-parity projectors for $p$-wave,
$$\chi_{\nu}^{(^1 P_1)}=\gamma_5 \phi^5_{\nu}$$
$$\chi_{\nu}^{(^3 P_2)}=\phi_{\nu\lambda}\gamma^{\lambda}, \quad
\chi_{\nu}^{(^3 P_1)}
=-\gamma_5[ \phi_{\nu} 
          - (\phi_{\lambda}\gamma^{\lambda})(\gamma_{\nu}-v_{\nu})],
\quad
\chi_{\nu}^{(^3 P_0)}=-(\gamma_{\nu}-v_{\nu}),
$$
we arrange them into a pair of doublets corresponding with two distinct
states of the total light angular momentum.
Thus the doublet of higher spin $(1^+, 2^+)$ is
$$\chi_{\nu}^{(\Uparrow)}
 = \left ( \begin{array}{c} 
           -\phi_{\nu\lambda}\gamma^{\lambda} \\
 \displaystyle
           \gamma_5 
       [ \varphi_{\nu} - \frac{1}{3}(\varphi_{\lambda}\gamma^{\lambda})
                                        (\gamma_{\nu}-v_{\nu})]
          \end{array} \right ), $$
and the lower spin doublet $(0^+, 1^+)$ is
$$\chi_{\nu}^{(\Downarrow)}
= \left (
   \begin{array}{c} 
    \gamma_5(\frac{1}{3}\varphi_{\lambda}
            -\frac{1}{\sqrt{2}}\phi_{\lambda})\gamma^{\lambda}  \\ 
    \frac{1}{\sqrt{3}} 
   \end{array}   \right ) (\gamma_{\nu}-v_{\nu}), $$
where $\varphi_{\nu}\equiv\phi^5_{\nu}+\frac{1}{\sqrt{2}}\phi_{\nu}$ and 
the constraint, $\gamma^{\nu}\chi_{\nu}^{(\Uparrow)} =0$ applies to the
components of $(1^+, \; 2^+)$  doublets. In fact when we take the trace 
according to Eq.(10) with the spin-parity projector of $(0^+, \; 1^+)$, 
the vector factor $(\gamma_\nu-v_\nu)$ in the $\chi_{\nu}^{(\Downarrow)}$ 
can be absorbed into $M^\nu(v,p)$, leaving us a scalar matrix in Dirac 
space. The ensuing analysis is thus exactly the same as the 
$(0^-, \; 1^-)$ doublet and we do 
not repeat it here. With respect to the projector for the 
$\chi_\nu^{(\Uparrow)}$ doublet, we construct
$$(M^{\nu})_{\alpha}^{\beta}=p^{\nu} [\bar \chi(p)]^{\sigma}_{\rho}\;
   [{\cal M}_0(p, v)]_{\sigma\alpha}^{\rho\beta}, $$
for $L'=0$ and like before, the scalar ${\cal M}_0$ can be expressed in 
terms of two unknown functions. When $L'=1$, we have  
$$(M^{\nu})_{\alpha}^{\beta}
 =p^{\nu}[\bar \chi_{\mu}(p)]^{\sigma}_{\rho}
 \;[{\cal M}_1^{\mu}(p, v)]_{\sigma\alpha}^{\rho\beta},$$
which formally has the the same structure as the $M$ in Eq.(13). 
In addition to those four form factors, we need two more to describe
$$(M^{\nu})_{\alpha}^{\beta}=[\bar \chi^\nu (p)]^{\sigma}_{\rho}
 \;[{\cal M}_1(p, v)]_{\sigma\alpha}^{\rho\beta},$$
making a total of six. A similar analysis for 
$L' \geq 2$ produces the forms
$$ (M^{\nu})_{\alpha}^{\beta}
 =p^{\nu}[\bar \chi_{(\mu_1\cdots\mu_{L'})}(p)]^{\sigma}_{\rho}
 \;[{\cal M}^{\mu_1\cdots\mu_{L'}}(p, v)]_{\sigma\alpha}^{\rho\beta}.$$
$$ (M^{\nu})_{\alpha}^{\beta}
 =g^{\mu_1\nu}[\bar \chi_{ ( \mu_1\mu_2\cdots\mu_{L'} ) }(p)]^{\sigma}_{\rho}
 \;[{\cal M}^{\mu_2\cdots\mu_{L'}}(p, v)]_{\sigma\alpha}^{\rho\beta}, $$
to each of which belong four form factors. Therefore we finish up with 
eight form factors. (As discussed before, there are four
unknown functions for the first $M^{\nu}$ above and another four for the 
second when $L'>1$ --- but only two when $L'=1$. However if $L'=0$, only 
the two form factors from the first $M^{\nu}$ contribute.) 
    
In summary, we have {\em two} universal form factors for the transition 
from heavy $p$-wave to light $s$-wave, {\em six} form factors to light 
$p$-wave, and {\em eight} to light $d$-wave or states of higher spin.

(3) $L$-wave $\rightarrow J'$

We are now in a position to complete our analysis for
the general case.  We rearrange the four states, $^3L_{L+1}, \;
^3L_{L}, \; ^3L_{L-1},$ and $^1L_{L}$ into a pair of doublets:
$$\chi_{ ( \nu_1\cdots\nu_L ) }^{(\Uparrow)}
 =\left ( \begin{array}{c} 
          -\phi^{L+1}_{ \{ \nu_1\cdots\nu_L\lambda \} }\gamma^{\lambda} 
          \\
\displaystyle
          \gamma_5 
   \left  [ \varphi^{L}_{ \{ \nu_1\cdots\nu_L \} }
            -\frac{1}{2L+1}\sum_k
             \varphi^{L}_{\{ \nu_1\cdots\bar k\cdots\nu_L\lambda\} }
             \gamma^{\lambda}(\gamma_{\nu_k}-v_{\nu_k}) \right ]
         \end{array} \right ), $$
and
$$\chi_{ ( \nu_1\cdots\nu_L ) }^{(\Downarrow)}
= \displaystyle   
  \sum_k      
\left ( \begin{array}{c} 
 \displaystyle   
    \gamma_5
\left (\frac{1}{2L+1}\varphi^{L}_{\{ \nu_1\cdots\bar k\cdots\nu_L\lambda\} }
           -\frac{1}{\sqrt{2}L}
            \phi^{L}_{\{ \nu_1\cdots\bar k\cdots\nu_L\lambda\} } \right )
            \gamma^{\lambda} 
          \\           
 \displaystyle   
     \frac{1}{L}\frac{\sqrt{2L\!-\!1}}{\sqrt{2L\!+\!1}}
     \left [ \phi^{L-1}_{\{ \nu_1\cdots\bar k\cdots\nu_L \} }
          \! -\! \frac{1}{2L\!-\!1}\sum_l
             \phi^{L-1}_{\{ \nu_1\cdots\bar k\bar l\cdots\nu_L\lambda\} }
             \gamma^{\lambda}(\gamma_{\nu_l}\!+\!v_{\nu_l}) \right ]
           \end{array} \right )\!(\gamma_{\nu_k}\!-\!v_{\nu_k}), $$
with $$\varphi_{ \{ \nu_1\cdots\nu_L \} }
=\phi^5_{ \{ \nu_1\cdots\nu_L \} }
 +\frac{1}{\sqrt{2}}\phi_{ \{ \nu_1\cdots\nu_L \} }.$$
(The `mixing angle' is uniform for all $L$.)
For the higher spin doublet $(L, L+1)$ we have the following 
$L'+1$-fold structures, when $L'\leq L$,
$$(M^{\nu_1\cdots\nu_L})_{\alpha}^{\beta}=p^{\nu_1}\cdots p^{\nu_L}
 [\bar \chi_{ ( \mu_1\cdots\mu_{L'} ) }(p)]^{\sigma}_{\rho}
\;[{\cal M}^{\mu_1\cdots\mu_{L'}}(p, v)]_{\sigma\alpha}^{\rho\beta},$$
$$(M^{\nu_1\cdots\nu_L})_{\alpha}^{\beta}
 =g^{\nu_1\mu_1}p^{\nu_2}\cdots p^{\nu_L}
 [\bar \chi_{ ( \mu_1\cdots\mu_{L'} ) }(p)]^{\sigma}_{\rho}
 \;[{\cal M}^{\mu_2\cdots\mu_{L'} }(p, v)]_{\sigma\alpha}^{\rho\beta}.$$
$$\cdots\cdots$$
and
$$ (M^{\nu_1\cdots\nu_L})_{\alpha}^{\beta}
 =g^{\nu_1\mu_1}\cdots g^{\nu_{L'}\mu_{L'}} p^{\nu_{L'+1}}\cdots p^{\nu_L}
 [\bar \chi_{ ( \mu_1\cdots\mu_{L'} ) }(p)]^{\sigma}_{\rho}
\;[{\cal M}(p, v)]_{\sigma\alpha}^{\rho\beta}.$$
As before to each of these objects corresponds four form factors, except 
the last one which just has two form factors. Altogether then there are 
$4L'+2$ form factors. On the other hand, if $L'> L$ it is easy to work 
out that the number of the form factors is $ 4(L+1)$. 
Turning next to the doublet of lower spin $(L-1, L)$, we can surely
absorb the factors like $ (\gamma_{\nu_k}-v_{\nu_k})$ into the overlap 
integral and thereby consider the case of a Lorentz tensor of rank $L-1$.
This leads us to the conclusion that the number of independent form 
factors is $4L'+2$ if $L' \leq L-1$, but $4L$ if $L'> L-1$.  

The situation for heavy to heavy transitions is simpler. Since the 
spin-parity projector for the final resonance is also a Rarita-Schwinger
object ($\gamma^\mu P_\{{\mu\cdots\}} = 0$), form factors 
$G_3^{(^{2S+1}L'_{J'})}$ and $G_4^{(^{2S+1}L'_{J'})}$ do 
not contribute, and  $G_1^{(^{2S+1}L^{\prime}_{J'})}$ and 
$G_2^{(^{2S+1}L'_{J'})}$ collapse into one form factor for an on-shell heavy
quark ($\not p = M'$).

\section{Exclusive Decay Rates}
Based on the matrix elements obtained previously, we shall now
evaluate rates for various exclusive processes including semileptonic, 
non-leptonic and rare decays. We shall restrict ourselves to 
pseudoscalar $\bar B$-decays into the light resonances of spin-$J$. Thus 
we make the substitutions $J\rightarrow0$ and $J'\rightarrow J$ in the 
earlier formulae. As expected, all of these exclusive rates are written 
in terms of 4 form factors. Here are the results case by case.

\subsection{Non-leptonic $B$-decays into $K$-meson resonances plus 
   charmonia}

We consider the 2-body hadronic decays $\bar B\rightarrow K^i(\bar c c),$
in which charmonia $(\bar c c)$ could be $J/\psi, \psi', \chi_{1c}, 
\eta_c,  {\sl etc}.$. The effective Hamiltonian relevant to processes 
$b\rightarrow s \bar c c$ is given by \cite{DTP}
\begin{equation}
H_{eff}=C\bar s\gamma^{\mu}(1-\gamma_5)b\;
         \bar c\gamma_{\mu} (1-\gamma_5)c, 
\end{equation}
with $$C=\frac{G_F}{\sqrt{2}}(c_1+c_2/3)V^*_{cs}V_{cb}.$$
Here $V_{cs}$ and $V_{cb}$ are elements of CKM matrix and $c_1$ and $c_2$ are 
Wilson coefficients.
Assuming factorization and using decay constants 
$f_Pq_{\mu}=<0|\bar c \gamma_{\mu}\gamma_5 c|(\bar c c)_P>$ 
for the pseudoscalar charmonium like $\eta_c$, and 
$f_V\phi_{\mu}=<0|\bar c \gamma_{\mu} c|(\bar c c)_V>$ 
for the vector charmonium like $J/\psi$, results in 
\begin{equation}
 H_{eff}(b\rightarrow s (\bar c c)_P)
 = Cf_P q_{\mu}\bar s \gamma^{\mu} (1-\gamma_5) b, 
\end{equation}
and
\begin{equation}
 H_{eff}(b\rightarrow s (\bar c c)_V)
 =Cf_V \phi_{\mu}\bar s \gamma^{\mu} (1-\gamma_5) b. 
\end{equation}

The exclusive decay rate for $\bar B\rightarrow K^i (\bar c c)_P$ 
depends only upon the longitudinal part of the matrix element, namely the
$F_0$-term of the parametrization in Eq.(3). Making use of the general
formula for polarization sums in the Appendix, we arrive at the decay rate
\begin{equation}
 \Gamma(\bar B\rightarrow K^i (\bar c c)_P)
=\frac{2^J(J!)^2  f_P^2}{8\pi (2J)!}|C|^2 
(\frac{M_i}{M^2})(M^2-M_i^2)^2
 \left [ \frac{(v\cdot p)^2}{M_i^2} - 1 \right ]^{J+\frac{1}{2}}
 |M^J F_0^{(J)}(v\cdot p)|^2,  
\end{equation}
with $v\cdot p =(M^2 + M_i^2 - M_P^)/2M.$ This longitudinal form
factor $F_0$ may be related our to our set $G_i$ via
$$\begin{array}{rl}
F_0^{(J)} 
=\frac{2}{M^{J-\frac{1}{2}}(M^2-M_i^2)}&
 [(M-v\cdot p)G_1^{(^{2S+1}L_J)} 
 \mp (M_i^2-M v\cdot p)G_2^{(^{2S+1}L_J)} \\
&\pm M G_3^{(^{2S+1}L_J)}- M_i^2 G_4^{(^{2S+1}L_J)}],
\end{array}$$
where the upper sign applies to $J=L$ and the lower one to $J=L\pm 1$. 

Similarly, the Lorentz invariant form factors in Eq.(2) for $V-A$ currents 
are related to our $G_i$ by
$$\begin{array}{l}
a_+^{(J)} 
=\frac{1}{M^{J-\frac{1}{2}}}
 (\frac{G_1^{(^{2S+1}L_J)}}{M}\pm G_2^{(^{2S+1}L_J)} - G_4^{(^{2S+1}L_J)})
 \\ \\
a_-^{(J)}   
=\frac{1}{M^{J-\frac{1}{2}}}
 (\frac{G_1^{(^{2S+1}L_J)}}{M}\mp G_2^{(^{2S+1}L_J)} + G_4^{(^{2S+1}L_J)})
 \\ \\
f^{(J)} 
=\frac{2}{M^{J-\frac{3}{2}}}
 [\pm G_3^{(^{2S+1}L_J)} +(v\cdot p) G_4^{(^{2S+1}L_J)}] 
\\ \\
g^{(J)} = \frac{2}{M^{J-\frac{1}{2}}}G_4^{(^{2S+1}L_J)}.
\end{array} $$
When $({\bar c}c)$ is a vector, the form factor $a_-^{(J)}$ does 
not contribute to the amplitude because $\phi_{\mu}q^{\mu}=0$. 
It is straightforward, though tedious, to  calculate the decay rate using 
Eq.(2); the result reads
\begin{equation}\begin{array}{rl}\displaystyle
  \Gamma(\bar B\rightarrow K^i (\bar c c)_V)
=& \displaystyle
  \frac{2^J(J!)^2 f_V^2}{16\pi (2J)!}|C|^2 
  \left (\frac{M_i^3}{M^2} \right )
  \left [ \frac{(v \cdot p)^2}{M_i^2} - 1 \right ]^{J-\frac{1}{2}}
 \\ \\ 
&\displaystyle 
 \left\{ \frac{8M^2}{M_V^2}\left |[\frac{(v \cdot p)^2}{M_i^2}-1]M^Ja_+^{(J)} 
        +\frac{1}{2}(\frac{v\cdot p}{M_i}-\frac{M_i}{M})
         \frac{M^{J-1}f^{(J)}}{M_i} \right|^2
\right .\\ \\ 
&\displaystyle \left .
        +\frac{J+1}{J} \left ( 
          \left | \frac{M^{J-1}f^{(J)}}{M_i} \right |^2 
          +[\frac{(v \cdot p)^2}{M_i^2}-1]\left |M^J g^{(J)} \right |^2 
 \right ) \right \},
\end{array}\end{equation}
where all form factors are fixed at $v\cdot p=(M^2+M_i^2-M_V^2)/2M.$

\subsection{Semileptonic $B$-decays into light flavoured meson resonances}

The amplitude for $\bar B\rightarrow X_q l \bar \nu_l$ contains hadronic
and leptonic currents
\begin{equation}
\frac{G_F}{\sqrt{2}}V_{qb}L_{\mu}<X_q(p^{\prime})|
\bar q \gamma^{\mu} (1-\gamma_5) b |\bar B(p)>, 
\end{equation}
in which $L_{\mu}=\bar u_l \gamma_{\mu} (1-\gamma_5)v_{\nu_l}.$ Once again 
$a_-^{(J)}$-term in Eq.(2) makes no contribution in the limit of massless
leptons and the differential distribution of the $\bar B$-decay is 
given by
\begin{equation}\begin{array}{rl}\displaystyle
\frac{d\Gamma}{dq^2 d\Omega_ld\Omega'}
&= \displaystyle
  \frac{2^J(J!)^2}{(4\pi)^5 (2J)!}
  \left ( \frac{G_F}{\sqrt{2}} \right )^2 |V_{qb}|^2 
  (\frac{M_i}{M})(\frac{q^2}{M^2})
  \left ( \frac{\Delta}{M_i} \right )^{2J-1}
\\ \\ 
&\displaystyle 
\left \{ \frac{8}{q^2M_i^2} [ \Delta^2-(k\cdot p)^2 ]  
\left |\Delta\,a_+^{(J)}+\frac{p'\cdot q}{2\Delta}f^{(J)}\right|^2
\right . \\ \\
&\left . \displaystyle
+\frac{J+1}{J} \left [ [\Delta^2+(k\cdot p)^2 ] 
 \left (\left |\frac{f^{(J)}}{\Delta} \right |^2 + |g^{(J)}|^2 \right ) 
                         +4(k\cdot p){\rm Re}(f^{(J)}\, g^{(J)*})
                  \right ]
 \right \},
\end{array}\end{equation}
where $\Omega_l$ is the solid angle of the
lepton in the rest ($l \bar\nu_l$) frame where $\vec q =0$, 
$\Omega'$ the angle of the final meson in the rest frame of
the initial meson and $ \Delta^2=M^2[ (v\cdot p)^2 - M_i^2 ]$.
As these form factors depend only on $q^2$, we integrate
over all angles to obtain
\begin{equation}\begin{array}{rl}\displaystyle
\frac{d\Gamma}{dq^2}
&= \displaystyle
  \frac{2^J(J!)^2}{48\pi^3 (2J)!}
  \left ( \frac{G_F}{\sqrt{2}} \right )^2 |V_{qb}|^2 
  (\frac{M_i}{M})^3 q^2
  \left ( \frac{\Delta}{M_i} \right )^{2J+1}
\\ \\ 
&\displaystyle 
\left [ \frac{4}{q^2M_i^2}   
\left |\Delta\, a_+^{(J)}+\frac{p^{\prime}\cdot q}{2\Delta}f^{(J)}\right|^2
+\frac{J+1}{J} \left (
\left |\frac{f^{(J)}}{\Delta} \right |^2 + |g^{(J)}|^2 \right ) 
\right ].
\end{array}\end{equation}
We notice that the contribution of $a_+^{(J)}$ is suppressed near the zero 
recoil point, $v\cdot p =M_i$, where the heavy quark approximation
is supposed to work well; the differential distribution is then
dominated by form factors $G_3^{(^{2S+1}L_J)}$ and 
$G_4^{(^{2S+1}L_J)}$.

\subsection{Radiative rare $B$-decays into $K$-meson resonances}

The radiative rare decays induced by the FCNC at the loop level of the 
standard model are mediated by an effective Hamiltonian \cite{C7}
\begin{equation}
H_{eff}=C_\gamma m_b\epsilon^*_\mu \bar s\sigma^{\mu\nu}q_\nu(1+\gamma_5)b,
\end{equation}
where $m_b$ is the mass of the bottom quark, $\epsilon^*_{\mu}$ the photon 
polarization vector, 
$$C_{\gamma}=-\frac{G_F}{\sqrt{2}}(\frac{e}{4\pi^2})C_7V^*_{ts}V_{tb},$$
and $C_7$ is a Wilson coefficient. The exclusive decay rate here reads
\begin{equation}
 \Gamma(\bar B\rightarrow K^i \gamma)
=\frac{2^J(J!)^2}{8\pi (2J)!}\frac{J+1}{J}|C_{\gamma}|^2 (m_bM)^2 
 \left [ \frac{(v\cdot p)^2}{M_i^2} - 1 \right ]^J  |H^{(J)}(v\cdot p)|^2, 
\end{equation}
with $v\cdot p = (M^2 + M_i^2)/2M.$ There is but a single form factor 
$H^{(J)}$ and this is related to our form factors $G_i,$ 
$$H^{(J)}=\pm\frac{G_3^{(^{2S+1}L_J)}}{M}+G_4^{(^{2S+1}L_J)},$$
with the upper (lower) sign for $J=L\quad (J=L\pm 1)$.
The ground state version of similar relation was obtained in ref. 
\cite{IW42}.

\section{Conclusion}

We have outlined the general Lorentz structure for matrix elements of 
current operators between meson states with arbitrary spins with 
particular focus on a pseudoscalar (or scalar) meson decaying
into resonances of higher spin. The matrix element for these processes
resembles very closely the extensively studied $0^-\rightarrow 1^-$
decays. Without reference to parity, {\em three} form factors pertain to 
the transverse (conserved) part of the current. One extra form
factor is needed to describe the longitudinal part. If the full angular 
distribution of the exclusive rate can be determined experimentally then
it is possible in principle to extract each of the 4 form factors.

Using the heavy quark approximation for the decaying heavy flavoured 
meson, we may achieve a great simplification in the matrix elements, 
reflected in a decrease of the number of form factors; such matrix elements
may be expressed in terms of a set of universal form factors which are 
independent of the mass and spin of the heavy quark inside the decaying
heavy meson, as well as the Dirac structure of the current operator. 
Four of these form factors, for instance, are sufficient to parametrize
any matrix elements between $0^-$ and spin-$J$ states. 
Importantly, this allows us to link various decay processes induced by 
different currents: for example, semileptonic decays via $V-A$ and rare 
decays by an effective current (arising from one-loop diagrams).

At the phenomenological level, we have formulated rates for various 
exclusive $\bar B$-decays into light resonances of higher spin and 
expressed them in terms of certain universal form factors. As more
experimental measurements of $\bar B$-decays become available in the
near future, we may hope to determine these form factors through different 
decay modes and thereby test the heavy quark approximation. To be sure 
all of these results can be applied to $D$-meson decays too, as long 
as we assume the charm quark is heavy enough compared to the QCD scale. 
In this way one may use the heavy flavour symmetry to relate universal 
form factors between $B$-meson and $D$-meson decays.  

A parallel analysis for baryonic states is currently being undertaken.

\acknowledgements

The authors are grateful to the Australian Research Council for their
financial support, under grant number A69231484.

\appendix{}
\section*{Polarization Sums}

We first carry out the polarization sums in the rest frame of the spin $J$
meson $\vec p'=0$. Representing such a polarization tensor by
a space-like, symmetric, traceless vector with $J$ indices, 
$\phi^{(\lambda)}_{i_1\cdots i_J}$, the fundamental formula is
$$ \sum_{(\lambda)} q^{i_1}\cdots q^{i_J} \phi^{(\lambda)}_{i_1\cdots i_J}
  \phi^{*(\lambda)}_{j_1\cdots j_J} q'^{j_1}\cdots q'^{j_J}
 =\frac{2^J (J!)^2 }{(2J)!} (|\vec q| |\vec q'|)^J P_J(\vec n\cdot \vec n')$$
where $n$ and $n'$ are unit vectors along arbitray vectors $\vec q$ 
and $\vec q'$.
Setting $q = q' = p$ we obtain the elementary result,
$$\displaystyle\sum_{\lambda}
  |p^{i_1}\cdots p^{i_J} \phi^{(\lambda)}_{i_1\cdots i_J}|^2
= \frac{2^J (J!)^2}{(2J)!} |\vec p|^{2J}. $$
By differentiating the first formula with respect to $q$ and $q'$ we may 
peel off indices, one at a time. Doing this just once for $q$ and $q'$, we get
$$ \begin{array}{l}\displaystyle\sum_{\lambda} 
 q^{i_2}\cdots q^{i_J}\phi^{(\lambda)}_{i i_2\cdots i_J}
 q'^{j_2}\cdots q'^{j_J}\phi^{(\lambda)*}_{j j_2\cdots j_J} \\
=\displaystyle
 \frac{2^J(J!)^2 }{J^2(2J)!} (|\vec q||\vec q'|)^{J-1} \left[\delta_{ij}P'_J
 -(n_in_j + n'_in'_j)P''_{J-1} + n_in'_j(P_{J-2}''-2P_{J-1}'')
 +n'_in_jP''_J\right ].
\end{array} $$
Then setting $q=q'=p$, one obtains   
$$ \begin{array}{l}\displaystyle\sum_{\lambda} 
 p^{i_2}\cdots p^{i_J}\phi^{(\lambda)}_{i i_2\cdots i_J}
 \phi^{(\lambda)*}_{j j_2\cdots j_J}p'^{j_2}\cdots p'^{j_J}\\
 =\displaystyle
 \frac{2^J (J!)^2}{(2J)!} |\vec p|^{2(J-1)} \left[
 \frac{J+1}{2J}\delta_{ij}+\frac{J-1}{2J}\frac{p_i p_j}{|\vec p|^2}\right].
 \end{array} $$
This trick can be continued to free up all the indices, but fortunately this
will not be required in what follows.

Boosting the above results to an arbitrary frame is easy: one simply makes the
replacements,
$$\delta_{ij}\rightarrow -g_{\mu\nu}+\frac{p'_{\mu}p'_{\nu}}{M^{\prime 2}}
 \equiv d_{\mu\nu}(p'),\quad
 p_i\rightarrow p_{\mu}
             -\frac{p\cdot p'}{M^{\prime 2}}p'_{\mu}, \quad
 \phi^{(\lambda)}_{i_1\cdots i_J} \rightarrow
 \phi^{(\lambda)}_{\mu_1\cdots\mu_J},$$
leading to
$$\displaystyle
\sum_{\lambda} |p^{\mu_1}\cdots p^{\mu_J}
 \phi^{(\lambda)}_{\mu_1\cdots \mu_J}(p^{\prime})|^2 =
\frac{2^J (J!)^2}{(2J)!} \left ( \frac{\Delta}{M'} \right )^{2J},$$
and
$$\begin{array}{l}\displaystyle
\sum_{\lambda} p^{\mu_2}\cdots p^{\mu_J}
 \phi^{(\lambda)}_{\mu\mu_2\cdots \mu_J}(p')p^{\nu_2}\cdots p^{\nu_J}
 \phi^{(\lambda)*}_{\nu\nu_2\cdots \nu_J}(p') \\ 
=\displaystyle
\frac{2^J (J!)^2}{(2J)!} \left ( \frac{\Delta}{M^{\prime}} \right )^{2(J-1)} 
\left [ \frac{(J+1)}{2J} d_{\mu\nu}(p') + \frac{(J-1)}{2J}
 \frac{(p_{\mu}-\frac{p\cdot p'}{M^{\prime 2}}p'_{\mu})
       (p_{\nu}-\frac{p\cdot p'}{M^{\prime 2}}p'_{\nu})}
      {( \frac{\Delta}{M'} )^{2} } \right ],
\end{array}$$
where $\Delta^2 \equiv p^4 + p'^4 + q^4 - 2p^2q^2 - 2p'^2q^2 - 2p^2p'^2$ 
and $q=p-p'$.


\begin{references}
\bibitem{fn1} In this and the following section $p$ ({\em not} $p'$) 
 denotes the  momentum of the final resonance and $v$ is the velocity of 
 the initial meson, which is a slight change of notation.
\end{references}
\end{document}